\documentstyle[emulateapj,epsf]{article} 

\def\reference{\par\noindent\hangindent=1cm\hangafter=1}

\newcommand{\eq}{\begin{equation}}
\newcommand{\ee}{\end{equation}}

\def\gapp{\ \lower 3pt\hbox{${\buildrel > \over \sim}$}\ }
\def\lapp{\ \lower 3pt\hbox{${\buildrel < \over \sim}$}\ }

\lefthead{Richer et al.}
\righthead{Old White Dwarfs}

\begin{document}
\submitted{Accepted for publication in the Astrophysical Jpournal}

\title{ISOCHRONES AND LUMINOSITY FUNCTIONS FOR OLD WHITE DWARFS}

\author {Harvey B. Richer$^{1,2}$, Brad Hansen$^3$, Marco Limongi$^1$,
Alessandro Chieffi$^4$, Oscar Straniero$^5$}
\affil {and}

\affil {Gregory G. Fahlman$^2$}

\vspace{0.5in}

\affil  {$^1$Osservatorio Astronomico di Roma}
\affil  {Via Frascati 33, Monteporzio Catone 00040, Italy}

\vspace{0.2in}

\affil  {$^2$Department of Physics and Astronomy}
\affil  {University of British Columbia, Vancouver, BC, Canada V6T 1Z1}

\vspace{0.2in}

\affil  {$^3$Canadian Institute for Theoretical Astrophysics}
\affil  {University of Toronto, 60 St. George Street, Toronto, ON, Canada
        M5S 3H8}

\vspace{0.2in}
 
\affil  {$^4$Istituto di Astrofisica Spaziale - CNR}
\affil  {C.P. 67, I - 00044, Frascati, Italy}

\vspace{0.2in}

\affil  {$^5$Osservatorio Astronomico di Collurania, Teramo, Italy}

\vspace{0.2in}

\begin{abstract}

Using a new grid of models of cooling white dwarfs, we calculate isochrones
and luminosity functions in the Johnson-Kron/Cousins and HST filter sets 
for systems containing old white dwarfs. These new
models incorporate a non-grey atmosphere which is necessary to properly describe the
effects of molecular opacity at the cool temperatures of old white dwarfs. 
The various functions calculated and extensively tabulated and plotted 
are meant to be as
utilitarian as possible for observers so all results are
listed in quantities that observers will obtain. The tables and plots
developed should eventually
prove critical in interpreting the results of HST's Advanced Camera observations
of the oldest white dwarfs in nearby globular clusters, in understanding the
results of searches for old white dwarfs in the Galactic halo, and in
determining ages for star clusters of all ages using white dwarfs.
As a practical application we demonstrate the use of these results
by deriving the white dwarf cooling age of the old Galactic cluster M67.
 
\end{abstract}

\keywords{white dwarfs --- globular clusters --- dark matter -- ages}

\section{Introduction}

The search for and use of old white dwarfs in determining the ages and star
formation histories in stellar systems was given an important lift recently
with the publication of new sets of models of cooling white dwarfs (Hansen
1998, 1999; Saumon and Jacobson 1999). These models included non-grey atmospheres
which are critical in understanding the luminosity and emergent spectrum of
white dwarfs whose temperatures fall below 4000K. It is the atmosphere which regulates
changes in the white dwarf's largely isothermal core and hence its cooling
time. Also, the behavior of the atmosphere is strongly dependent on its
composition, particularly the amount of hydrogen and helium, as helium does
not form molecules whereas hydrogen does at cool temperatures. Hydrogen
molecules thus provide
a dramatic opacity source which must be including in the modelling in order to understand the emergent flux from the star. Thus careful teatment of the
physics is essential to properly interpret the luminosities and colors of old white dwarfs.

As an added incentive, the recent microlensing results in the direction of the 
Magellanic Clouds (Alcock {\it et al.} 1997a, 1997b; Renault {\it et al.} 1997) suggest
that a sizeable fraction (perhaps half or even larger) of the dark matter in
the Galactic halo could be tied up in stellar objects with masses
near $0.5 M_{\odot}$. This suggests old white dwarfs as the likely candidate
although other possibilities remain (neutron stars, primordial black holes or
other exotica). Although all these candidates for the Galactic dark matter have
their problems, numerous searches are now underway to attempt to locate these
objects and there already exists a few possible old white dwarf candidates in the
Hubble Deep
Field (Ibata {\it et al.} 1999) and in the general field of the Galaxy (Harris {\it et al.} 1999). All the searches for old white dwarfs
must be guided by appropriate
cooling models, isochrones and luminosity functions. In this paper we present
all the above as an aid in directing these endeavors. The compilations here
are much more extensive than those in Chabrier 1999 and use a different set of models, those of Hansen. The tables and
plots are all in the observers plane, in the Johnson-Kron/Cousins
$VRI$ color system or the HST system and attempts have been made to make the data as utilitarian for observers
as possible. For this reason we have presented not just the cooling models but
have developed white dwarf isochrones and luminosity functions both for clusters
and for the Galactic halo. These are
the quantities which will actually be observed when the
Advanced Camera for Surveys on HST eventually penetrates to the faint end of the
white dwarf cooling sequence in a globular cluster or when a wide area
ground-based survey detects a sizeable sample of old halo white dwarfs. 

 \section{The White Dwarf Cooling Models}

The white dwarf models presented here are based on the code of Hansen and
Phinney 1998. The addition of new atmospheric models (Hansen 1998, 1999) has
led to a revision of the cooling ages and observational appearance of old
white dwarfs and it is these that we will use.
The only models we present are those for C-O cores (without separation energy)
and with hydrogen-rich atmospheres.
Chemical separation may lengthen the ages slightly (Salaris {\it et al.} 1997; Hansen 1999).
As such, the omission of this contribution represents the most conservative assumption, that is 
the fastest cooling.
It is only for hydrogen-rich 
atmospheres that the white dwarfs
remain bright (brighter than $M_V = 18$) for times comparable to the Hubble 
time. This is due to the strong
opacity of molecular hydrogen. The helium-rich models, which do not possess
this
opacity source, cool much more rapidly and become fainter than $M_V = 
18$ on a timescale less than 6 Gyr. Because of the strongly non-blackbody
colors of cool hydrogen-rich white dwarfs caused by the hydrogen molecules redistributing the
emergent flux, the stars actually become bluer in $(V - I)$
when $T_{eff}$ drops below about 3500K. At this point there is little change
in $M_V$ as the stars evolve but the $V$$R$$I$ colors become very different from those of black bodies {\it of any temperature} and these colors
may be a key to the discovery of old white dwarfs. 

In Figure 1 we illustrate the effect that the added $H_2$ opacity has on the
observed cooling track of a $0.5M_{\odot}$ C-O core, H-rich (DA) white dwarf.

 Here
we compare the cooling sequence for such a model by Hansen with an extrapolated model using
Wood's 1992, 1995 interiors and Bergeron {\it et al.'s} 1995 atmospheres. Down to an
$M_V$ of about 16 (age slightly older than 7 Gyr for both models), both sets
of models agree quite well, but as the star gets cooler and more molecules
are able to form, the effect of the molecular opacity increases and 
the two models differ
enormously. {\it In $(V-I)$, old white dwarfs are blue not red}.

Table 1 contains these new cooling models in
Johnson/Kron-Cousins $VRI$ and Table 2 in the HST filters while Figure 2 plots
the $M_V$, $(V-I)$ cooling sequences from Table 1. The colors can be quite different in these two different filter sets 
as the strong $H_2$ opacity produces sharp flux peaks in the emergent white
dwarf spectrum.

\centerline{{\vbox{\epsfxsize=8cm\epsfbox{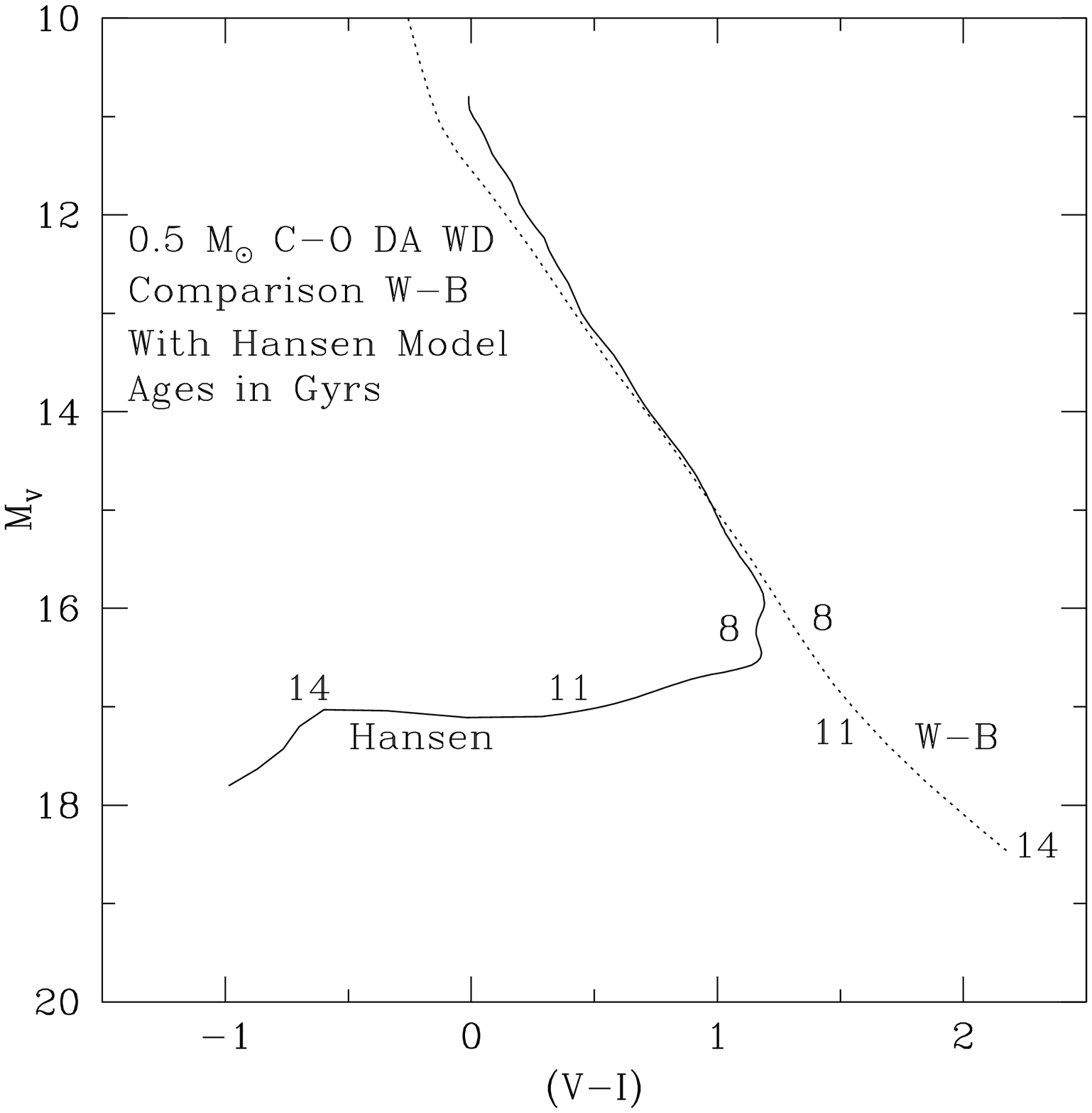}}}}
\noindent{\small
 Fig. 1: New $0.5 M_{\odot}$ white dwarf cooling model of Hansen 1998, 1999 compared
with a similar mass model constructed from the interiors of Wood 1992 
and Bergeron {\it et al.} 1995 atmospheres (W-B). The main differences set in at around 8 Gyr where the effects of atmospheric $H_2$ opacity become important.
\label{figTen1}
}
\vspace{2mm}

\noindent The colors measured by the observer then depend critically 
on the positions 
of the
transmission peaks of the filters. The HST colors
 are calculated using the Holtzmann {\it et al.} 1995 bandpasses and
 the transformations they use to express fluxes in $V$, $R$ and $I$.

\centerline{{\vbox{\epsfxsize=8cm\epsfbox{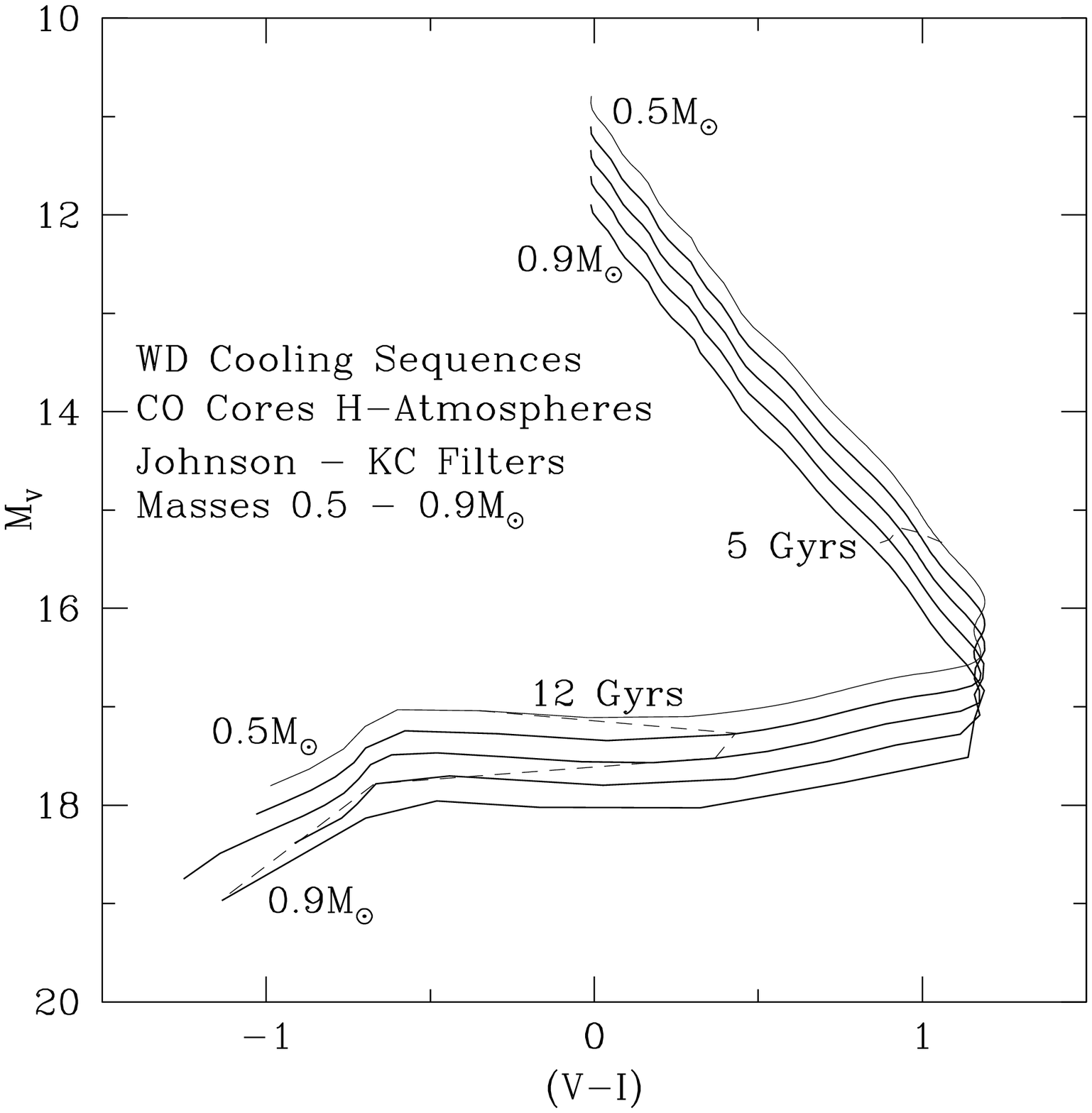}}}}
\noindent{\small
Fig. 2: Cooling sequences for C-O core, hydrogen-rich white dwarfs of
varying mass. The curves shown are for Johnson-Kron/Cousins filters. Constant
ages of 5 and 12 Gyr are indicated on the diagram.

\label{figTen2}
}
\vspace{2mm}

The mass range in the models 
varies from $0.5 - 0.9 M_{\odot}$ and the sequences all begin at about an age of 0.2 Gyr as it takes this
long for the white dwarfs to lose the imprint of their initial conditions. The
models terminate when $T_{eff}$ drops below about 2000K which is the limit of the
opacity tables used.

Figure 3 plots the color-color diagram in the Johnson/Kron-Cousins
photometric system for a white dwarf of mass $0.7M_{\odot}$ with ages indicated
along the sequence. The wild deviations from black body colors
are evident in this diagram as the oldest and coolest white dwarfs get
dramatically bluer in the $(V-I)$ color and somewhat bluer in $(V-R)$.

\centerline{{\vbox{\epsfxsize=8cm\epsfbox{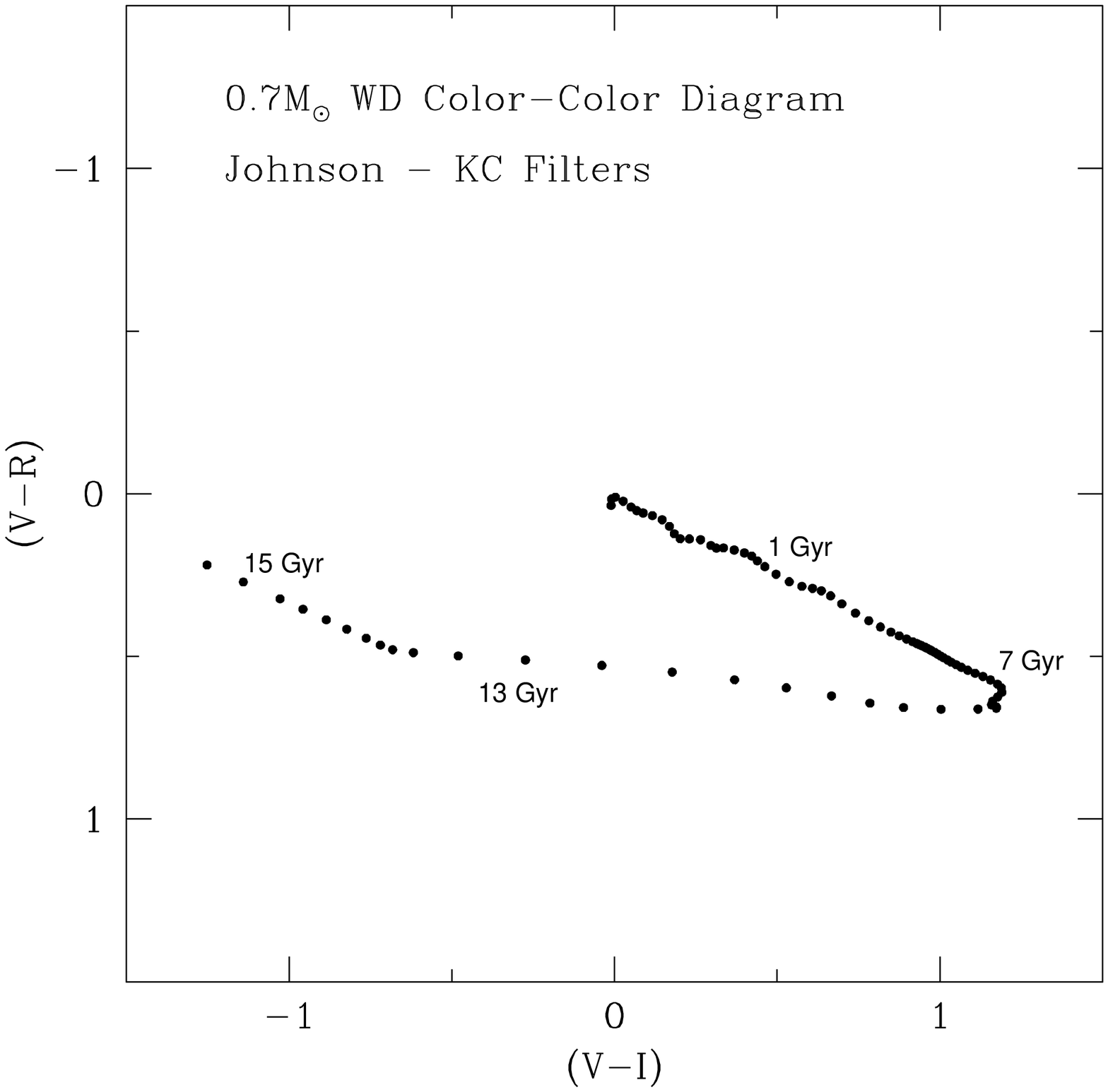}}}}
\noindent{\small
Fig. 3: ($V-R$), ($V-I$) color-color diagram for $0.7M_{\odot}$ hydrogen-rich
white dwarfs in Johnson-Kron/Cousins filters. When $H_2$ opacity becomes
important for ages older than about 8 Gyr, the colors deviate strongly from
those of black bodies.

\label{figTen3}
}

\section{White Dwarf Isochrones}

When white dwarfs are observed in an open or globular cluster it is not
strictly correct to compare their location in the cluster color-magnitude
diagram with a theoretical cooling sequence of some mass as has generally
been done in the past  
(e.g. Richer {\it et al.} 1995, 1997, 1998; Cool {\it et al.} 1996; Renzini {\it et al.} 1996). The reason
for this is that the oldest white dwarfs in these clusters have evolved 
from the most massive stars originally in the cluster (up to the maximum 
mass that produces white dwarfs),
and because more massive progenitors produce more massive remnants,
the older white dwarfs should be more massive. This has generally not been a
problem with the clusters observed thus far as the range in white dwarf masses
has been relatively small, and, in any case, the initial mass function of clusters
is expected to yield
many fewer massive stars and hence few massive white dwarfs. However, when
large ground-based telescopes or HST eventually penetrate to the termination point of the white
dwarf cooling sequence in a globular cluster, and thus cover a wide range
in white dwarf masses, it will be extremely important to have white dwarf
isochrones ready to interpret the data as opposed to just cooling sequences
for some mass.

For these reasons we have calculated isochrones for white dwarfs in 
star clusters
with a wide range in age. All the isochrones shown are derived from solar metallicity models.
The isochrones were constructed by (a) beginning with a
white dwarf mass of $0.9M_{\odot}$, the maximum mass model that we had
available, and 
(b) using an
initial-final mass relation constructed from Herwig's 1995 data at the high mass
end mated to the results from Gibson {\it et al.} 1999 for M67 and M4
at the low mass end to determine the 
mass of the main sequence progenitor. In using these data we are mixing
metal rich and metal poor relations, however, at the moment this is all
that can be done if we wish to stick with empirical results. (c) The stellar
evolutionary models of Dominguez {\it et al.} 1999 were then employed 
to determine the lifetime
of the main sequence star ($A_{ms}$) up to the end of the AGB. (d) The age of the white dwarf is
then simply $T_{iso} - A_{ms}$ where $T_{iso}$ is the age of the isochrone that we are calculating.
(e) The absolute magnitude and color of the white dwarf of interest was then 
obtained from the cooling model for a white dwarf of mass $0.9 M_{\odot}$ and age  $T_{iso}$ - $A_{ms}$. (f) We then decremented the mass, interpolating within
the models, and repeated the
process until a white dwarf mass of $0.5M_{\odot}$ was reached (the minimum
white dwarf mass model available), at which point
the calculations were halted. 

Figure 4 illustrates these isochrones for a range of ages likely to
be of interest in any application. The hook to the blue
in the isochrones for ages less
than about 7 Gyrs is not due to the effects of $H_2$ opacity (these stars are
too
hot for $H_2$ to form) but is caused by the fact that the white dwarfs at
the bottoms of these curves come from massive main sequence stars. These 
produce heavier white dwarfs which follow cooling sequences which lie below
those of lighter degenerates (more massive white dwarfs have
smaller radii and are thus less luminous at a given temperature). It is only
for ages older than 8 Gyrs where the effect of the $H_2$ opacity is seen. 

\centerline{{\vbox{\epsfxsize=8cm\epsfbox{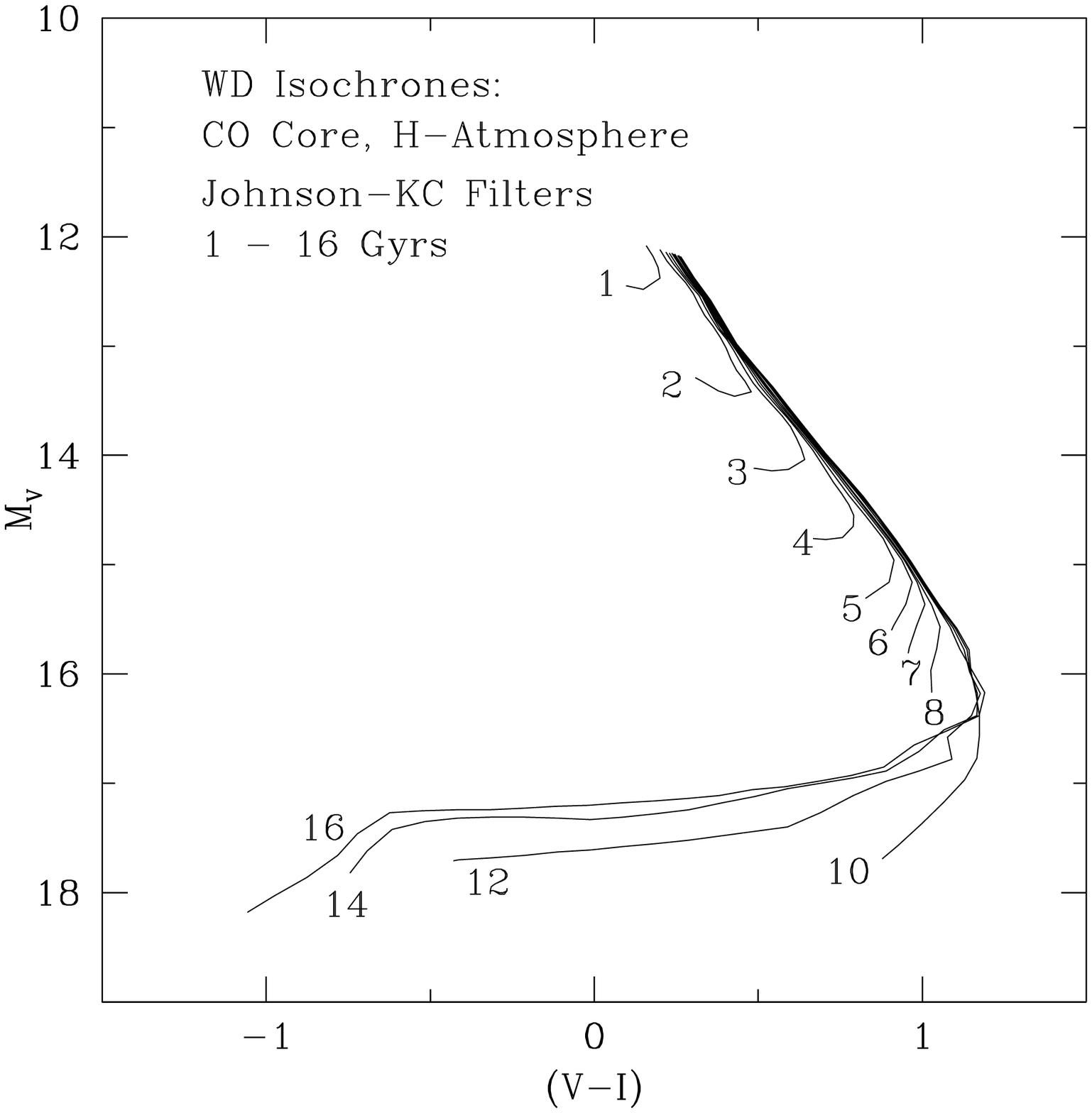}}}}
\noindent{\small
Fig. 4:  White dwarf isochrones in Johnson-Kron/Cousins filters. Isochrone 
ages are
indicated.
\label{figTen4}
}
\vspace{2mm}

Figure
5 illustrates some detail for the 11 Gyr isochrone, showing the mass of the
white dwarf itself and that of its main sequence precursor. 
Tables 3 and 4 list selected isochrones in both the 
Johnson-Kron/Cousins system and in the HST filters. The columns in these tables
are the white dwarf mass ($M_{\odot}$), the mass of the progenitor 
($M_{\odot}$) and $T_{eff}$, $M_V$, ($V-R$) and ($V-I$) of the white dwarf.
Details of the initial-final mass relation that we used can be obtained from these Tables.

\centerline{{\vbox{\epsfxsize=8cm\epsfbox{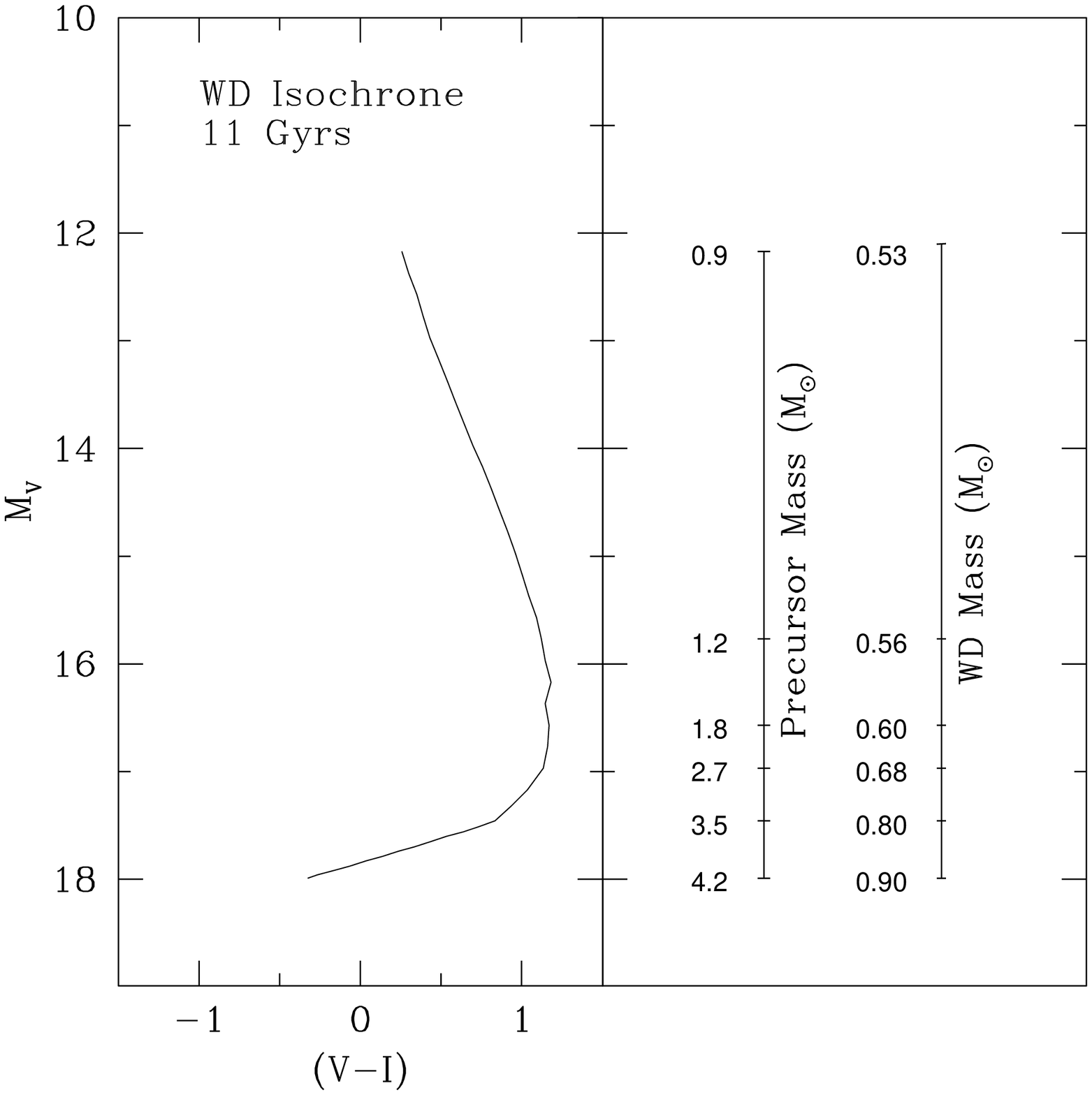}}}}
\noindent{\small
Fig. 5: Details of the 11 Gyr white dwarf isochrone indicating white dwarf
and precursor masses.  
\label{figTen5}
}

\section{Cluster White Dwarf Luminosity Functions}

The white dwarf luminosity function in a star cluster, in the absence of
dynamical evolution, contains information
on the initial mass function (IMF) and age of the cluster. In fact, it
will eventually be possible to use the white dwarf luminosity function in
an old star cluster (e.g. a globular cluster) to
extend the observed main sequence mass function up to massive stars that 
many billions of
years ago evolved in to white dwarfs (Richer {\it et al.} 1997).

White dwarf luminosity functions for different ages were constructed in the following manner. 
(a) The isochrones and the
initial-final mass relation were used to set the maximum and minimum main sequence masses for a cluster of a particular age.
(b) Based on the IMF used, a random extraction of a main sequence mass in this
range was
made and this then yielded a white dwarf mass from the initial-final mass relation. (c) From the isochrones, the $M_V$ of this white dwarf was then
obtained. (d) This was repeated 1,000 times and eventually renormalized to 100
white dwarfs for each cluster.

Figure 6 illustrates such
luminosity functions for a Salpeter IMF ($n(m) \propto m^{-\alpha}$ where $\alpha = 2.35$, solid line) and a much flatter IMF ($\alpha = 1.3$, dashed line) 
which is more
in line with the steepest IMFs being found at the low mass end in globular clusters (Piotto and
Zoccali 1999). The main
feature to note in this diagram is the manner in which the peak of the cluster
white dwarf luminosity function marches toward lower luminosity as the cluster age increases.
This is then a potentially powerful technique for determining cluster ages
that is largely independent of isochrone fitting to the turn off region
of a cluster. As can be seen, the effect of even a radical change in the IMF slope has a rather
small influence on the morphology of the white dwarf luminosity function and it
appears that this is unlikely to be a sensitive method of investigating cluster
IMFs. For this reason we only tabulate functions for Salpeter IMFs.

\centerline{{\vbox{\epsfxsize=8cm\epsfbox{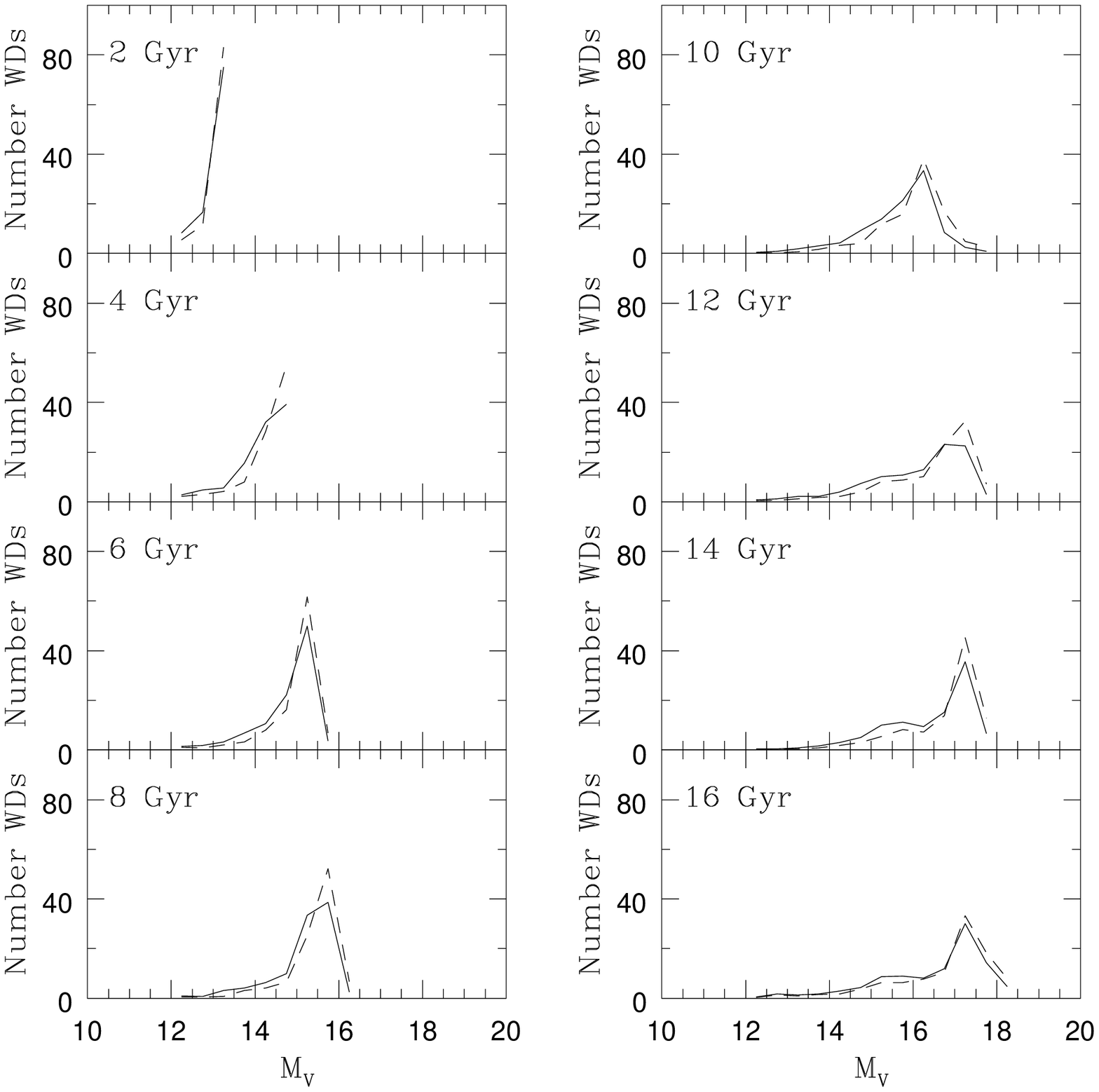}}}}
\noindent{\small
Fig. 6:  Differential luminosity functions for white dwarfs in clusters of varying ages.
The solid lines are for Salpeter IMFs ($\alpha = 2.35$) while the dashed lines
are for significantly flatter IMFs ($\alpha = 1.3$).

\label{figTen6}
}

\vspace{2mm}

These luminosity functions are listed in Table 5 for those in 
Johnson-Kron/Cousins filters and in Table 6 for those calculated in the HST
filter set. In these tables the number of white dwarfs is normalized to 100
and the columns are, respectively, the absolute $V$ magnitude of the middle
of the bin, the number of white dwarfs in that bin, the cumulative number of
white dwarfs, the mean mass of the white dwarfs and of the progenitors.

As a last point regarding white dwarf luminosity functions in clusters, we inquire whether
information about the age of a cluster can be obtained if the turnover 
in the white dwarf luminosity function is {\it not} observed, but only if a bright
portion (e.g. to $M_V = 15$) is seen. This of course has potential practical
applications as the termination points of white dwarf sequences will only be
possible to observe in the nearest globular clusters even with HST and the Advanced Camera for Surveys. To investigate this we superimpose in Figure 7 synthetic white dwarf luminosity functions for 10, 12 and 14 Gyr old clusters.
The numbers of white dwarfs indicated are those expected from a single WFPC2
field at 6 core radii from the center of the globular cluster M4. If the functions in Figure 7 are compared only 
down to $M_V = 15$, it becomes clear that virtually no useful information is 
obtained regarding the age of the cluster.
The turnover in the luminosity function must be
observed in order to constrain the cluster age.

\section{The White Dwarf Cooling Age of M67}

In an earlier paper Richer {\it et al.} 1998 presented and discussed the
observed white dwarf luminosity function in the open cluster M67 which
has a turn off age of about 4 Gyr (Montgomery {\it et al.} 1993). Here we 
compare this function with synthetic luminosity functions in order to derive
the white dwarf cooling age of the cluster. In the previous paper we did 
not have
access to such synthetic functions so the current derivation of the cluster cooling age will supercede the results in the earlier paper.

\centerline{{\vbox{\epsfxsize=8cm\epsfbox{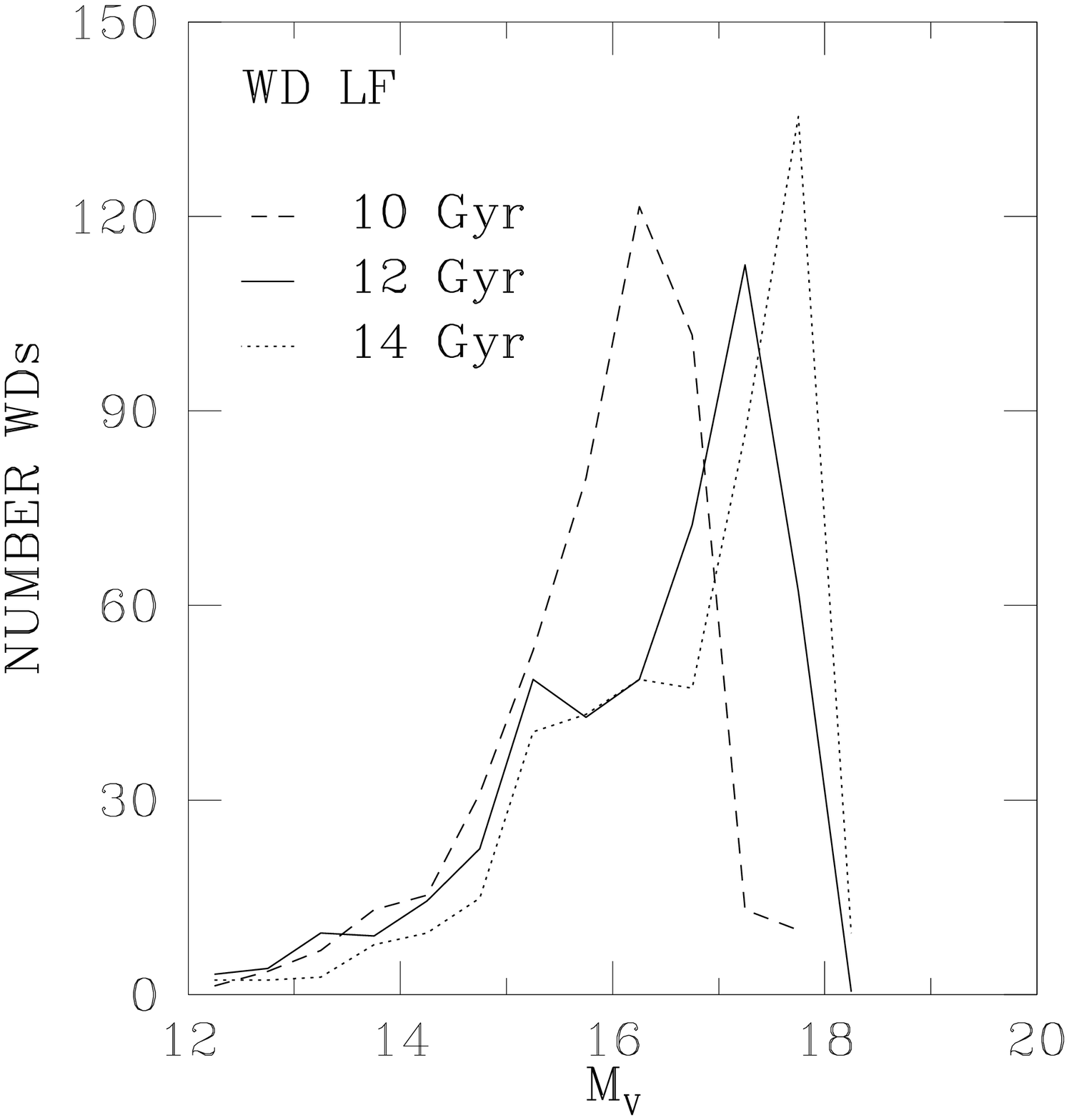}}}}
\noindent{\small
Fig. 7: Synthetic white dwarf luminosity functions in the HST filters
for clusters with ages of 10, 12 and 14 Gyr. The
numbers of white dwarfs are those expected in the Galactic globular cluster M4
in a single WFPC2 field at 6 core radii from the cluster center (see Richer {\it et al.} 1997).

\label{figTen7}
}

\vspace{2mm}

Figure 8 displays the observed cumulative white dwarf luminosity function in 
M67
from Richer {\it et
al.} 1998 (heavy solid line) compared with synthetic luminosity functions for clusters with ages of 
3, 4 and 5 Gyrs. The synthetic 

\centerline{{\vbox{\epsfxsize=8.0cm\epsfbox{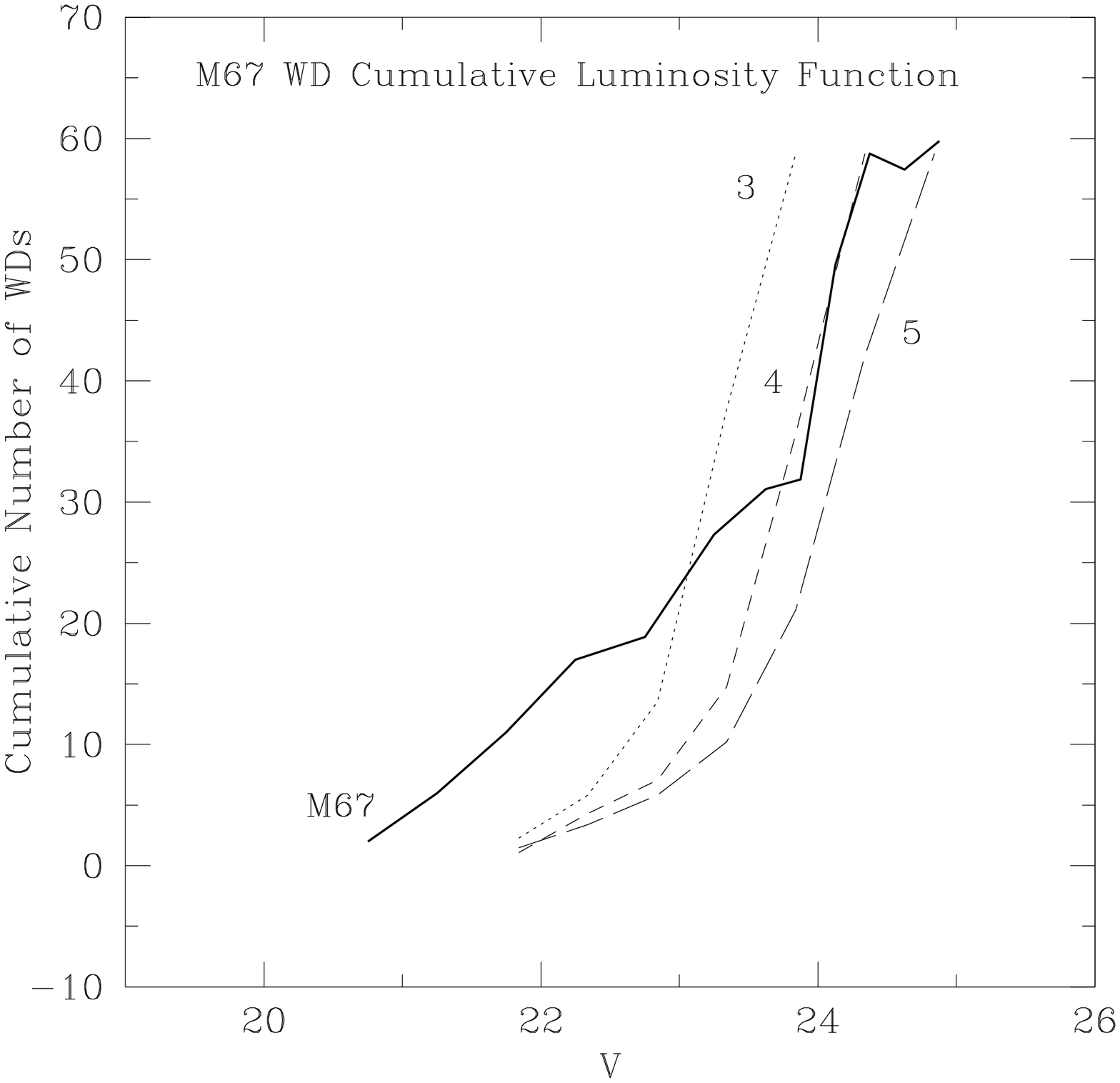}}}}
\noindent{\small
Fig. 8: The cumulative M67 white dwarf luminosity function (heavy solid line)
compared with synthetic functions of ages 3, 4 and 5 Gyr, all with Salpeter
IMFs. The location of the break in the M67 luminosity function and the general
fit to the synthetic functions suggest a white dwarf cooling age for M67 
 near 4 Gyr. 

\label{figTen8}
}

\vspace{2mm}

\noindent functions have been shifted so as to
represent a cluster with an apparent $V$ distance modulus of 9.59 (Montgomery 
{\it et al.} 1993) and they have been normalized to contain the same number of white dwarfs
that are observed in the cluster (58).  All
the synthetic functions have Salpeter IMFs 
but the actual choice of the IMF,
within rather broad limits, has rather little effect on the final results as
could be deduced from Figure 6. 
From the faint end of the luminosity function seen in M67, 
it is clear that the cooling 
age of the cluster is larger than 
3 Gyr, less than 5 Gyr and that the 4 Gyr synthetic luminosity function is an excellent
match to the luminosity function of the faintest observed cluster white dwarfs.

This result indicates that a properly constructed synthetic white dwarf
luminosity function compared with data should be a robust and reliable age
indicator, and that it will be an important tool in establishing ages for
old clusters in the Galaxy.

We note in passing that the observed white dwarf luminosity function in M67 has
a well populated tail of stars to high luminosity, many more stars than are
predicted by the models. Varying the IMF, even by a rather large amount, could
not make the fit of the synthetic function to the observations significantly
better as the precursor mass range among the bright M67 white dwarfs is quite small. The origin of this tail is
not currently understood but might be related to the high binary fraction in
the cluster (see Richer {\it et al.} 1998 for further discussion) or to 
some deficiency
in the cooling models which {\it overestimates} the true rate of cooling of young
white dwarfs. If the binary scenario is correct, the excess number of bright white dwarfs could be produced by 
making a relatively large number of helium core white dwarfs via 
truncated stellar
evolution. Such objects fade less rapidly than C-O white dwarfs as they have
a greater heat capacity per unit mass.

\section{The White Dwarf Luminosity Functions in the Galactic Halo}

The microlensing experiments in the direction of the LMC
seem to be indicating  
that $60 \pm 20$\% of the dark matter in the Galactic halo is tied
up in $0.5^{+0.3}_{-0.2} M_{\odot}$ objects (Alcock {\it et al.} 1997a, 1997b; Renault {\it et al.} 1997). This naturally suggests old white
dwarfs although other possibilities exist 
(e.g. neutron stars, primordial black holes). The possibility that white
dwarfs are important contributors to the Galactic dark matter has been
considered for some time now (Larson 1986; Silk 1991; Carr 1994), but with the
microlensing results this scenario has taken on increased viability.

Chabrier 1999, Chabrier {\it et al.} 1996, Gibson and Mould 1997, and Adams and Laughlin 1996 have all pointed out that
if indeed this scenario is correct, the IMF of the white dwarf precursors could
not have had a Salpeter form but might have been more Gaussian in shape and
peaked
near $2.7M_{\odot}$. For this reason we have calculated halo luminosity functions for
both Salpeter IMFs and those of the form $\Phi(m) = \exp^{-({\overline m/m)}^{\beta_1}}m^{-\beta_2}$ with $\overline m  = 2.7$, $\beta_1 = 2.2$ and $\beta_2 = 5.75$ (Chabrier {\it et al.} 1996).

Under this scenario, old white dwarfs will be plentiful in the Galactic halo but difficult to
detect because of their intrinsic faintness. The local number of such objects
can be determined simply from the local dark matter density 
($0.0079 M_{\odot}$/pc$^3$) (Alcock {\it et al.} 1997a; Chabrier and M{\'{e}}ra 1997; Gould, Flynn and Bahcall 1996) and the mean white dwarf mass ($\left\langle M_{WD}\right\rangle$)
through

\begin{displaymath} {Local~Number~WDs~/pc{^3}} = {\frac{0.0079} {\left\langle M_{wd}\right\rangle}} .
\end{displaymath}

A synthetic halo white dwarf luminosity function for a given age ($T_{field}$)
and limiting magnitude ($V_{lim}$) was calculated as follows. 
First, from the cluster luminosity function for an age $T_{field}$, we obtained the maximum and minimum main sequence masses as before as well as 
the brightest white dwarf ($M_V^{min}$) in the cluster. We then set the distance $R_{max}$
out to which we would fill a volume
with white dwarfs as

\begin{displaymath} {R_{max} = 10^\frac {(V_{lim} - M_{V}^{min} + 5)}{5}} .
\end{displaymath}

\noindent From the IMF we then extracted a main sequence star of a given mass
and determined
the associated white dwarf mass $M_{wd}$ through the initial-final mass relation. This
white dwarf was then placed randomly at a distance of $R_{wd}$ inside the volume so as to keep the density constant. From the isochrone we then obtained the $M_V$ of this
object and obtained its apparent $V$ magnitude from 

\begin{displaymath} {V = M_V - 5 +5~log~R_{wd}} .
\end{displaymath}

\noindent The average mass of the white dwarfs in the volume came simply 
from

\begin{displaymath} {\left\langle M_{wd}\right\rangle  =  \sum \frac {M_{wd}}{N}} ,
\end{displaymath}

\noindent and the total number of white dwarfs observed in the volume would then be

\begin{displaymath} {N_{tot} = \frac {0.0079}{\left\langle M_{wd}\right\rangle} * \frac {4}{3}\pi R^3_{max}} .
\end{displaymath}

\noindent This was done $N$ times until $N \ge N_{tot}$ at which point the
calculations were terminated.

In this way we constructed synthetic halo luminosity functions including
all the stars in the mass range allowed by the models for ages of 10, 14 and 16 Gyrs. The volume size was chosen so that all the
stars brighter than $V = 28$ would be counted. This limiting magnitude
constitutes a
reasonably faint limit but not so faint that the halo density variation would be
important for luminosity functions constructed with Chabrier {\it et al.} 1996
IMFs. With $V_{lim} = 28$ and  $M_{V}^{min} =$ 14.5, 15.5 and 16.0 for ages of 10, 14 and 16 Gyr with a Chabrier IMF, 
$R_{max}$ was 5.0, 3.2 and 2.5 kpc respectively.
However,  when we used a Salpeter IMF it was 
12.5 kpc for all ages. For the Chabrier IMFs
the number of stars inserted in to the volume was $\sim 1.6 \times 10^{5}$ per
square degree of field for the 10 Gyr halo ($\sim 2.0 \times 10^4$ for 16 Gyr) while
it was $\sim 2.5 \times 10^6$ for the Salpeter IMF at all 3 ages. 

We can only 
reasonably 
calculate halo luminosity functions for stars that are very local
(so as to keep the density constant and not to have the functions depend on the
direction of viewing) so this is a reasonable
assumption for Chabrier-type IMFs (about a 6\% error is made by not including
the $R^{-2}$ halo density variation for a 16 Gyr halo for viewing towards the Galactic poles), but not very good for Salpeter functions 
(looking in the same direction, an overestimate
of about a factor of 4 in the counts from the most distant region of the volume results from ignoring the density fall off). This is not a critical point as,
if indeed the MACHOs are old white dwarfs, their
precursors are unlikely to have been formed with a Salpeter IMF.
In any case, the Salpeter IMF counts exceed those with Chabrier IMFs by
a much larger amount than that caused by density variations. For example,
 for a 14 Gyr halo and a one square degree field to $V = 27.5$, the
cumulative counts with HST filters for a Salpeter IMF are 34,000 (assuming constant density)
to be compared with 1,210 for those with
a Chabrier IMF.
The average white dwarf mass for a halo of this age  
with a Chabrier IMF was $0.62M_{\odot}$
while it was $0.57M_{\odot}$ for a Salpeter function.

Tables 7 and 8 contain halo white dwarf luminosity functions for Johnson-Kron/Cousins and HST
filter systems, respectively, for ages of 10, 14 and 16 Gyr all 
with Chabrier IMFs.
For reasons outlined above, the functions
for Salpeter IMFs 
are not very accurate so they should be considered as
illustrative only. Even so, there are a number of interesting conclusions that
can be derived from the plots (Figure 9) and Tables 7 and 8. First, the
white dwarf luminosity functions based on a Salpeter IMF always predict many
more stars than do those for a Chabrier function. For example, the Salpeter-based
function for 14 Gyr predicts that 5 old white dwarfs should be found in the Hubble Deep
Field (HDF) to $V = 26.0$ whereas there is at most 1 possible candidate
(Ibata {\it et al.} 1999, Hansen 1999). This by itself seems capable of
excluding old white dwarfs formed with a Salpeter IMF as the entire source
of the Galactic dark matter. By contrast, using the luminosity function
generated with a 
Chabrier IMF suggests that only 0.2 should be found. This model
is clearly not as yet excluded by the data. 

\centerline{{\vbox{\epsfxsize=8cm\epsfbox{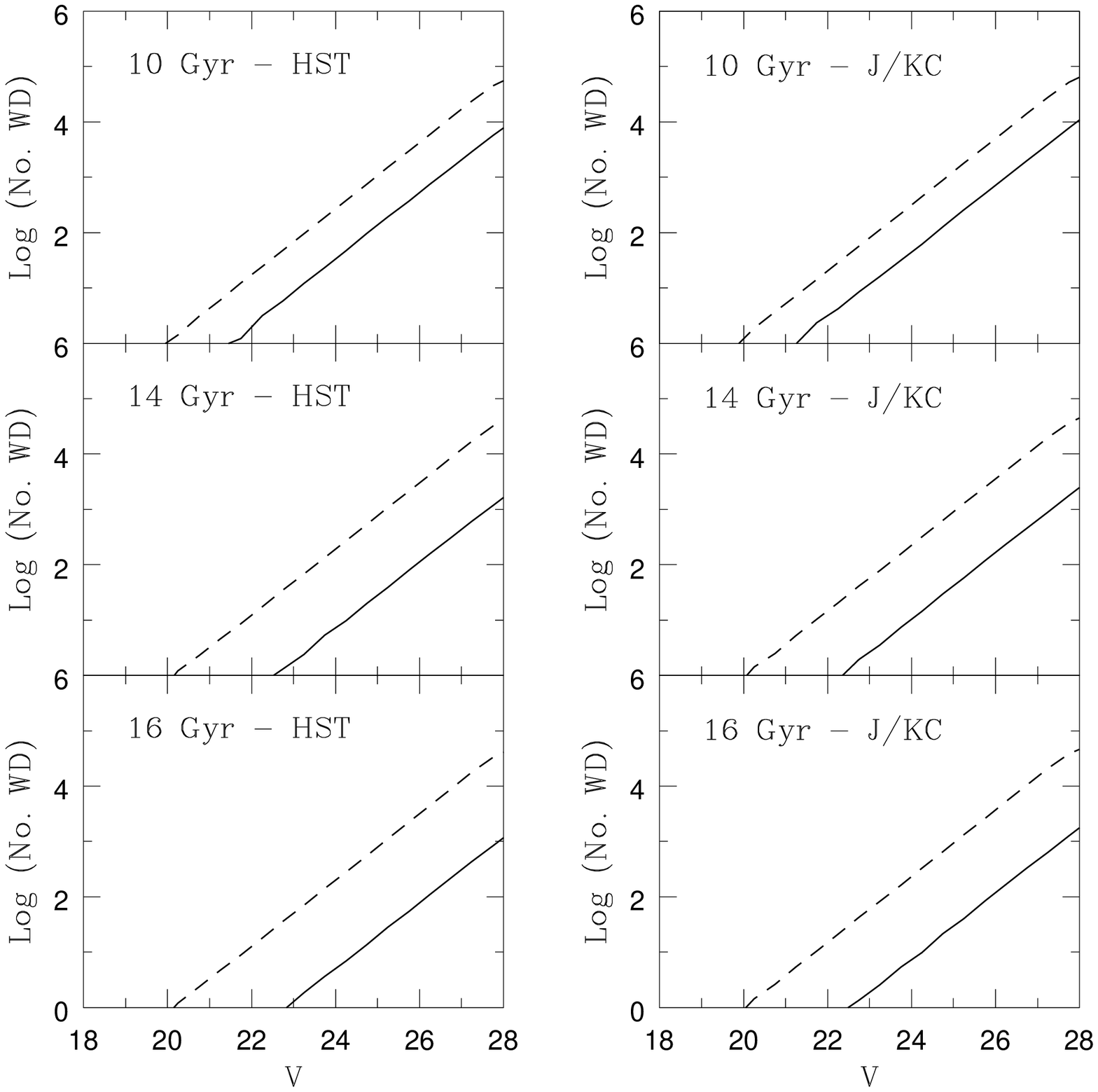}}}}
\noindent{\small
Fig. 9:  Cumulative white dwarf luminosity functions in the Galactic halo.
The y-axis is the logarithm of the number of stars per square degree under the assumption
that white dwarfs make up 100\% of the local dark 
matter density and that they all have hydrogen atmospheres.
 Solid lines are for a Chabrier 1999 type IMF while the
dashed line is for a Salpeter IMF.
\label{figTen9}
}
\vspace{2mm}

For a 10 Gyr old halo and a Chabrier IMF, to $V = 28.0$, we expect to find 14
old white dwarfs in the
HDF (if the 
dark halo is 100\% hydrogen-rich white dwarfs) but only 2 if 
it is as old as 16 Gyr. This, then, coupled with the colors of the stars, is a potentially powerful method of
establishing both the time of formation of the Galactic halo and possibly
shedding some light on the
nature of dark matter in the Galaxy. With the advent of the Advanced Camera for Surveys on HST, experiments of this sort covering larger areas than the
HDF will become
eminently feasible.

\acknowledgements
HBR thanks the director and staff at the Osservatorio Astronomico di Roma
for providing a delightful working environment
and support during the period in which this paper was developed and written. The research
of HBR and GGF is supported, in part, through grants from the Natural Sciences and Engineering Research Council of Canada.

\end{document}